\documentstyle[12pt]{article}

\def\be{\begin{equation}}
\def\ee{\end{equation}}
\def\ben{$$}

\def\een{$$}

\def\ba{\begin{array}{c}}
\def\ea{\end{array}}
\def\p{\partial}
\begin{document}

\titlepage
\vspace*{2cm}

 \begin{center}{\Large \bf
${\cal PT}-$symmetrized supersymmetric quantum
mechanics\footnote{Work supported by the grant Nr. A 1048004 of GA
AS CR.}}\end{center}

\vspace{10mm}

 \begin{center}
Miloslav Znojil\footnote{e-mail: znojil@ujf.cas.cz}

 \vspace{3mm}

\'{U}stav jadern\'e fyziky AV \v{C}R, 250 68 \v{R}e\v{z}, Czech
Republic

\end{center}

\vspace{5mm}

\section*{Abstract}

Supersymmetry between bosons and fermions is modeled  within
${\cal PT}-$symmetric quantum mechanics. A non-Hermitian
alternative to the Witten's supersymmetric quantum mechanics is
obtained.

\newpage

\section{Introduction}

Final chapters of the current textbooks on quantum mechanics
usually mention its natural extensions towards relativistic
kinematics and towards the quantization of fields (cf., e.g.,
\cite{Messiah}). Recently, fundamental concepts like symmetries
and interpretations have been re-considered within quantum
mechanics itself. For example, Witten \cite{Witten} proposed the
so called supersymmetric (SUSY) quantum mechanics (cf. its review
\cite{Khare}) and Bender and Boettcher \cite{BB} initiated a study
of an innovative formalism of ${\cal PT}$ symmetric quantum
mechanics (cf. also \cite{BBjmp}). In what follows, we intend to
discuss and further develop a certain overlap or combination of
the the latter two schemes in the direction indicated recently in
our letter \cite{PLB}.

Before moving to the deeper technical details, let us start from
an overall setting of the stage. Indeed, there exists a palpable
contrast between the well developed formalism of the traditional
Hermitian quantum mechanics of ref. \cite{Messiah} and the
striking incompleteness of interpretation of the alternative
non-Hermitian ${\cal PT}-$symmetric formalism of ref. \cite{BB}.
As mentioned by the referee of this paper, the gross mathematical
consistency of the former theory is already well understood since
H. Weyl. The essence of this theory is most clearly presented in
the language of the self-adjoint operators and/or their
essentially self-adjoint extensions \cite{Reed}. In contrast, the
unanswered open questions abound in the latter context. Even the
fundamental conjecture of the connection between the reality of
the spectrum and the related ``unbroken" ${\cal PT}$ symmetry is
just a more or less intuitive hypothesis at present. Many of its
illustrative manifestations are puzzling \cite{Mandal}. At the
same time, quite recently there emerged several indications that
its ``natural" formulation may be provided by resurgence theory
(cf. subject number 34M37 within the Mathematics Subject
Classification 2000 scheme) rather than by functional analysis
(ref. \cite{Delaba} can be recalled for an updated review of the
related literature).

Similarly, in the language of physics, the multi-faceted character
of the possible physical interpretations of the Witten's scheme
\cite{Witten} cannot be compared, at the present stage of
development at least, with the first-step nature of our fairly
simple-minded proposals in ref. \cite{PLB}. At the same time, it
is necessary to imagine that our latter representation of
supersymmetry ``lives" in an unusual space. In a way depending on
its explicit physical interpretation (which is very often being
related to the zero-dimensional field theory \cite{BM} and, hence,
need not necessarily coincide with the traditional versions of
quantum mechanics), one often speaks about the ``states" which are
manifestly regularized by their ${\cal PT}$ symmetrization. These
states need not necessarily survive in the appropriate Hermitian
limit. This is a crucial aspect of our forthcoming considerations
\cite{refer} and has to be kept in mind whenever one tries to
understand properly the relations between the non-equivalent
formalisms as depicted in the following diagram
 \ben
  \ba
  \\
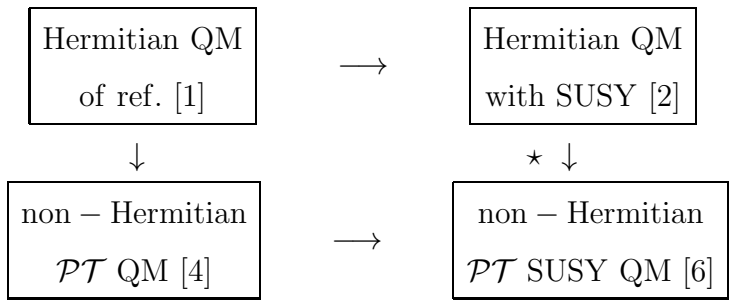
 \begin{array}{|c|}
 \hline
 {\rm Hermitian\ QM}\\
  {\rm  of\ ref.\ \cite{Messiah}}\\
 \hline
 \ea \ \ \ \ \ \ \
 \stackrel{}{ \longrightarrow }
  \ \ \ \ \ \ \
 \begin{array}{|c|}
 \hline
 {\rm Hermitian\ QM}\\
 {\rm with \ SUSY\ \cite{Witten}}\\
 \hline
 \ea
\\
 \ \ \downarrow  \  \ \ \ \ \ \
 \ \ \ \ \ \ \ \ \ \ \ \
 \ \ \ \ \ \ \ \ \ \ \ \
\ \ \ \ \  \star \ \downarrow \ {\rm }\  \
\\ \
 \begin{array}{|c|}
 \hline
 {\rm non-Hermitian}\\
 {\rm {\cal PT}\ QM\  \cite{BB}}\\
 \hline
 \ea \ \ \ \
 \ \
 \stackrel{}{ \longrightarrow }
 \ \ \
 \ \ \
  \begin{array}{|c|}
 \hline
 {\rm non-Hermitian}\\
  {\cal PT}\
 {\rm SUSY\  QM\ \cite{PLB}}\\
 \hline
\ea
\\
\\
\ea
 \een
In our forthcoming text we shall solely pay attention to the
$^\star -$marked correspondence. We shall outline a consistent
approach to supersymmetry after the weakening of the Hermiticity
to the mere ${\cal PT}$ symmetry.

For the sake of clarity, our considerations will be presented in
an elementary language inspired by the possible ${\cal PT}$
symmetrization of the Calogero two-particle solvable model
\cite{Calogero}. In combination with the methodical analysis of
the ${\cal PT}$ symmetric quartic anharmonic oscillators of ref.
\cite{PLB} this will enable us to modify slightly the current
interpretation of the ${\cal PT}$ symmetry itself. We shall also
analyze anew some of its possible consequences. This material will
be split in the sketchy description of the Witten's supersymmetry
in section 2, similar sketchy outline of the ${\cal PT}$ symmetry
in section 3 and their synthesis in section 4.

\section{Hermitian SUSY }

Energies $E^{(HO)}_n=2n+1$ of bound states generated by the
one-dimensional harmonic oscillator Hamiltonian $ H^{(HO)}=
p^2+x^2$ belong to the wave functions $\psi_n(x)$  with the
definite parity, ${\cal P}\psi_n(x)=\psi_n(-x) =(-1)^n\psi_n(x)$,
$n = 0, 1, \ldots$. Let us now employ this particular example and
recollect its re-interpretation within supersymmetric quantum
mechanics.

\subsection{Harmonic oscillator in supersymmetric picture}

In accord with the review of the supersymmetric quantum mechanics
by Cooper et al \cite{Khare} we can consider the left-shifted and
right-shifted harmonic oscillator Hamiltonians
 \ben
 H^{(L)}= p^2+x^2-1,
 \ \ \ \ \ \ \
 H^{(R)}= p^2+x^2+1
 \een
and notice their almost complete isospectrality,
 \ben
 E^{(L)}_m=2m, \ \ \ m = 0, 1, \ldots,
 \ \ \ \ \ \
 E^{(R)}_k=2k+2, \ \ \ k = 0, 1, \ldots.
 \een
Wave functions of the same energy become arranged in doublets,
 \be
 \psi^{(L)}_{n+1}(x)\ \longleftrightarrow \
\psi^{(R)}_{n}(x) , \ \ \ \ \ \ \ n= 0, 1, \ldots. \label{pairs}
 \ee
Formally this can be interpreted as a consequence of an underlying
supersymmetry $sl(1/1)$ since a representation of this algebra
working with both commutators and anticommutators can be generated
by the three two-by-two matrices one of which is defined in terms
of our two toy Hamiltonians,
 \ben
{\cal H}= \left [
 \begin{array}{cc}
H^{(L)}&0\\ 0&H^{(R)}\ea \right ] = \left [
 \begin{array}{cc}
  B^{} A^{}
&0\\ 0& A^{} B^{}  \ea \right ].
 \een
The well known factorization of the Hamiltonian is then defined in
terms of the harmonic oscillator superpotential $W^{(HO)}(x) = x$,
 \ben
A=A^{(HO)} = \frac{d}{dx} + W^{(HO)}(x), \ \ \ \ \ \ \ \ \ \ \
B=B^{(HO)} =- \frac{d}{dx} + W^{(HO)}(x).
 \een
The other two generators or ``supercharges" read
 \be
{\cal Q}=\left [
 \begin{array}{cc}
0&0\\A^{}&0 \ea \right ], \ \ \ \ \ \ \tilde{\cal Q}=\left [
 \begin{array}{cc}
0& B^{}
\\
0&0 \ea \right ]\  \label{key}
 \ee
and close the required superalgebra,
 \ben
 [ {\cal H},{\cal Q} ]=[ {\cal
H},\tilde{\cal Q} ]=0 , \ \ \ \ \ \ \{ {\cal Q},{\cal Q} \}= \{
\tilde{\cal Q},\tilde{\cal Q} \}=0, \ \ \ \ \ \
 \{ {\cal Q},\tilde{\cal Q}
\}={\cal H}.
  \een
Formally the whole scheme with its Riccati equation background
(cf. ref. \cite{Khare} for more details) is firmly rooted in the
nineteen century mathematics \cite{Darboux} and offers an
explanation of the exact solvability of many potentials after a
suitable change of the superpotential $W(x)$.

\subsection{The model of Calogero in supersymmetric picture}

In the present setting let us note that the supersymmetrized pairs
of states of the same energy are characterized by their parity,
 \ben
 {\cal P} \psi^{(L)}_{m}(x) =
-
 {\cal P} \psi^{(R)}_{k}(x), \ \ \ \ \ \ \ m = k+1,
 \ \ \ \ \ k = 0, 1, \ldots .
 \een
This feature finds an interesting interpretation within the so
called Calogero model of the two particles interacting via the
harmonic oscillator forces complemented by a short-term
``repulsion" \cite{Calogero}. The complete two-body Hamiltonian
reads
 \ben
 H^{(Cal)} = -
\frac{\p^2}{\p{x_1}^2} - \frac{\p^2}{\p{x_2}^2}
 + \left [
\frac{1}{2}   \,(x_1-x_2)^2
 +\frac{g}{
 (x_1-x_2)^{2}}\right ]\ .
  \een
After a routine elimination of the centre-or-mass coordinate
${R}=(x_1+x_2)/\sqrt{2}$ one is left with the radial
Schr\"{o}dinger equation
 \be
 \left [ -\frac{d^2}{d {r}^2} +
{r}^2+\frac{1}{2}\,\frac{g}{{r}^{2}}-E \right ] \psi({r}) = 0
\label{SE2}
 \ee
in the relative coordinate ${r}=(x_1-x_2)/\sqrt{2}$. This is a
singular ordinary differential equation and $E$ is the energy in
the centre-of-mass system. The singularity acquires the common
centrifugal form whenever we introduce the (in general,
non-integer) ``angular momentum"
 \be
\ell =\ell(g) = -\frac{1}{2} + \sqrt{ \frac{1}{4}+\frac{g}{2} } \
.\label{fu}
 \ee
The strongly repulsive centrifugal-like core can be perceived as
impenetrable. The normalizability of the wave function at $g \in
[3/2,\infty)$ implies that $ \psi^{(admissible)}({r}) $ vanishes
at $r=0$. By ``brute force", Calogero extended such an
impenetrability feature to all the couplings $g > -1/2$ by
demanding that it takes place for the weak repulsion (or, if
necessary, attraction) as well. The related explicit boundary
conditions
 \be
 \begin{array}{ll}
 \lim_{{r} \to 0} \, \psi^{(admissible)}({r}) = 0, \ \ \ \ \ \ &
 g \in ( 0, 3/2 )
 ,\\
 \lim_{{r} \to 0} \,[{r}^{-1/2} \psi^{(admissible)}({r})] = 0,
  \ \ \ \ \ \ &
 g \in (-1/2,0)
 \ea
 \label{re}
  \ee
were more thoroughly discussed in the related recent comment
\cite{Comment} containing the discussion of such a regularization
interpreted as a selection of the ``admissible" or ``physical"
solutions at any $g>-1/4$.

It may be useful \cite{refer} to re-emphasize here that the choice
of the regularization is fairly conventional, indeed. In the
various modern applications of the formalism of quantum mechanics,
the whole family of the ``nonstandard" models may prove equally
useful. The formal freedom of choosing any essentially
self-adjoint extension as our ``physical" Hamiltonian operator in
the interval of $g \in (-1/2, 3/2)$ enables us to modify
accordingly its desirable physical meaning \cite{Ditt}.

Under our present conventions we may summarize that the Calogero's
``non-communication rules" (\ref{re}) make the Calogero's equation
selfconsistently defined on both the half-axes ${r} \in
(0,\infty)$ and  ${r} \in (-\infty,0)$. In a way dictated by the
pure physics one can join these separate branches in both the
symmetric (= bosonic) and antisymmetric (= fermionic) manner,
 \be
 {\cal P}\psi^{(bosonic)}(x) = + \psi^{(bosonic)}(x),
 \ \ \ \ \
  {\cal P}\psi^{(fermionic)}(x) = - \psi^{(fermionic)}(x)
  \label{statistics}
 \ee
observing that

\begin{itemize}

\item
the operator ${\cal P}$ of parity can be re-interpreted as an
exchange of particles, within the framework of the sufficiently
non-harmonic Calogero model at least;

\item
tentatively, we can try to move to the smaller couplings $g
> -1/2$ and pick up the $g=0$ special case. Then, within the
above-mentioned supersymmetric arrangement (\ref{pairs}), we
obtain a new interpretation of supersymmetry as a very natural
mapping between the Calogerian bosons and
fermions~(\ref{statistics}).

\end{itemize}

\noindent We can add that the Hermitian supersymmetric
transformation unavoidably fails in all the $g \neq 0$ cases where
the superpotential $W(x)$ becomes singular. More details can be
found in chapter 12 of the review \cite{Khare}. Here we shall
propose a systematic remedy in a way inspired by our letter
\cite{PLB}.

\section{${\cal PT}$ symmetric Hamiltonians }

In the usual Hermitian conjugation $H = H^+$ the superscript $^+$
means the transposition $H \to H'$ combined with the complex
conjugation, $H^+= {\cal T} H'{\cal T}$. Within the ${\cal PT}$
symmetric quantum mechanics of ref. \cite{BBjmp} another type of
conjugation is necessary.

\subsection{Complexifications of smooth oscillators}

Let us briefly review the main features of the ${\cal PT}$
symmetric quantum mechanics. Its basic idea is simple and dates in
fact back to the old paper by Caliceti et al \cite{Caliceti}. In
this work the complex forces of a ``minimally" perturbed type
$V(x) = x^2 + i\,\lambda\,x^3$ have been shown to posses the real
spectrum at the sufficiently small and real couplings $\lambda$.
Using the symbols ${\cal P}$ (parity) and ${\cal T}$ (complex
conjugation) the probable relevance of the ${\cal PT}$ invariance
$V(x) = {\cal PT} V(x) {\cal PT}$ of this model has been
emphasized and re-emphasized in different contexts
\cite{DB,Alvarez}.

The recent explicit formulation of the hypothesis connecting the
reality of spectrum to the unbroken ${\cal PT}$ symmetry of wave
functions belongs to Bender et al \cite{BB,BBjmp}. It has been
verified on several different solvable examples.  During this
verification an important role has been played by the partially
solvable systems \cite{BBjpa,dekadic} and by the systematic ${\cal
PT}$ symmetrization of the so called shape invariant family of the
polynomially solvable models \cite{shapein,Eckart}. After a return
to their Hermitian limit, their classification is offered by the
supersymmetric quantum mechanics~\cite{Khare}.

A key to the relevance of the ${\cal PT}$ symmetrization can be
seen in the ambiguity of quantization resulting from a complex
deformation of the coordinate axis. In this way the analytic
potentials $V(x)$ admit a change of the spectrum caused by a mere
deformation of the integration path (i.e., asymptotic boundary
conditions) {\em in the complex plane}. This idea first appeared
in the context of field theory \cite{BM}. Its most elementary
illustrations have been mediated by the regular harmonic
oscillator \cite{BB} and by its partially solvable anharmonic
modifications of the sextic type \cite{BT}. A decisive progress in
the study of the ${\cal PT}$ symmetric systems of this type has
been achieved by Buslaev and Grecchi \cite{BG} who discovered an
unexpected byproduct of the elementary ${\cal PT}$ symmetric
complexifications which lies in a facilitated regularization of
the centrifugal singularities.

\subsection{${\cal PT}$ symmetric regularization }

In the spirit of paper \cite{BG} the range ${r} \in (0,\infty)$ of
many solvable radial Schr\"{o}doinger equations can be extended to
the whole complex contour $r=r(t)$ parameterized by a real $t \in
(-\infty,\infty)$. Quite often, the most elementary implementation
of the latter idea is represented by the real line shifted
slightly downwards,
 \ben
 r(t) = t - i\,\varepsilon, \ \ \ \ \ \varepsilon > 0.
 \een
One can further deform this curve $r(t)$ with $|r(t)| \to \infty$
for $t \to \pm \infty$ to many other suitable shapes. For the
harmonic oscillator example the curve only has to stay within the
asymptotic wedges in which the dominant part $\exp(-r^2/2)$ of our
explicit wave functions decreases, $|{\rm Im\ }r(t)| <|{\rm Re\
}r(t)|$, $ |t| \gg 1$ (cf. \cite{BB}).

Thorough repetition of a ${\cal PT}$ symmetric analysis of the
Calogerian singular example (\ref{SE2}) in \cite{hopt} (and of its
quartic \cite{quartic} and decadic \cite{dekadic} modifications)
revealed that a core of the difference between the regular and
regularized cases lies in the necessity of introduction of the
(say, upward-running) cut in our complex plane of $r$ in the
majority of cases \cite{Alvarez}. In the other words, the
quantization (and spectrum) can be influenced by the continuation
of $r(t)$ on the other Riemannian sheets. Such a type of the
subtle non-equivalence deserves a due care. Its exactly solvable
example was constructed by Cannata et al and can be found,
slightly hidden and implicit, in sec.~5 of their letter
\cite{Cannata}.

\section{${\cal PT}$ symmetric supersymmetry}

We are now prepared to find a way towards a synthesis of the
separate supersymmetric and ${\cal PT}$ symmetric modifications of
the traditional quantum mechanics. Our main idea is that after a
regularization of the centrifugal-like singularities many
difficulties encountered in the classical SUSY in connection with
the singular superpotentials could be in principle avoided via the
${\cal PT}$ symmetric regularization.  This returns us  once more
to the general experience with quantum theory where our choice or
specification of the physical meaning of the Hamiltonians can
remain, sometimes, quite flexible and adapted to our specific
phenomenological needs \cite{Ditt}.

\subsection{Anharmonic oscillators with the $x^{-2}$ core}

Teaching by example once more, let us pick up the first nontrivial
(viz., quartic anharmonic) example. In a way close to our present
discussion this example has been first studied by the purely
non-perturbative means by Flessas \cite{Flessas}. For the purposes
of inserting this example in the new supersymmetric scheme let us
also recall the results of the papers \cite{BBjpa} and
\cite{quartic}. We recollect that the potentials
 \be
V^{(sm)}(x)= -4i(x-i\varepsilon) -(x-i\varepsilon)^4, \label{pota}
 \ee
 \be
V^{(sp)}(x)= \frac{2}{(x+i\varepsilon)^2} -(x+i\varepsilon)^4
\label{potbe}
 \ee
possess the respective exact zero-energy solutions
 \ben
\psi^{(sm)}(x)= (x-i\varepsilon) \, \exp \left ( -i\,\frac{
(x-i\varepsilon)^3}{3} \right )\ \in \ L_2(-\infty,\infty),
 \een
 \ben
\psi^{(sp)}(x)= \frac{1}{x+i\varepsilon} \, \exp \left (
+i\,\frac{ (x+i\varepsilon)^3}{3} \right )\ \in \
L_2(-\infty,\infty).
 \een
We can, therefore, introduce the usual superpotentials
 \be
W^{(sp/sm)}(x)=-{ \left [ \frac{d}{dx}\psi^{(sp/sm)}(x) \right ]
/
\psi^{(sp/sm)}(x) } = \pm \left [
 \frac{1}{x\pm i\varepsilon}-
i\,(x\pm i\varepsilon)^2 \right ]\  .\label{susyp}
 \ee
In a way proposed in our recent study \cite{PLB} we employ the
standard definition of the supercharges (\ref{key}) containing the
nonstandard creation- and annihilation-type operators
 \ben
 A=A^{({\cal PT})}
 = {\cal T}\cdot \left [ \frac{d}{dx} + W^{(sm)}(x) \right ], \
 \ \ \ \ \ \ \ \ \ \ B=B^{({\cal PT})} =\left [ - \frac{d}{dx} +
 W^{(sm)}(x) \right ]\cdot {\cal T}.
 \een
This preserves all the underlying supersymmetric algebra $sl(1/1)$
and extends the concept of the supersymmetric partners to the case
of our non-Hermitian Hamiltonians.

\subsection{${\cal PT}$ symmetrized Calogero model at $A=2$}

All the Hermitian harmonic oscillators with a central symmetry
(i.e., real potentials $V(\vec{r})=|\vec{r}|^2$ in $D$ dimensions)
are described by the ordinary differential ``radial"
Schr\"{o}dinger equation of the form (\ref{SE2}) with $g/2
=\ell(\ell+1)$. This equation is solvable in terms of the
confluent hypergeometric functions and its integer of half-integer
parameter $\ell = (D-3)/2 + L $ depends on an integer quantum
number $L$ of the free angular motion. This implies that we can
write all the exact normalizable solutions in terms of Laguerre
polynomials in a way which is fully parallel to the well known
Hermitian case.

On a move to the singular and ${\cal PT}$ symmetrized Calogerian
harmonic oscillator (\ref{SE2}) with, in general, non-integer
$2\ell$ (and with the straight line $ r(t) = t - i\,\varepsilon$
for definiteness), we can recollect the results obtained in our
paper \cite{hopt}. There, the well known ``regular" harmonic
oscillator bound state solutions $\psi^{(reg)}(r)$ which behave
like $r^{\ell+1}$ near the origin were complemented by their
``irregular-like" partners $\psi^{(irreg)}(r) ={\cal
O}(r^{-\ell})$. The latter states remain formally acceptable
whenever $\varepsilon \neq 0$. At the same time, they just play a
purely formal role and ``naturally" disappear in the ``standard"
or ``physical" non-${\cal PT}$ Hermitian limit $\varepsilon \to
0$. Hence, no contradictory predictions can emerge. The physics
remains unchanged while merely its mathematical presentation is
modified. In fact the overall picture becomes simplified in a way
resembling the simplification of algebraic equations after one
moves from the real line to complex plane.

\begin{itemize}

\item
Our two sets of solutions form the non-equidistant spectrum $ \{
E^{(reg)}_n, E^{(irreg)}_n \}$ where
 \ben
 E^{(reg)}_n = 4n+2\ell+3,
\ \ \ \ \ \ \
 E^{(irreg)}_n =4n-2\ell+1,\ \ \ \ \ \ n = 0, 1, \ldots.
 \een
These ``energies" exhibit an $\alpha \leftrightarrow -\alpha$
symmetry with $\alpha = \ell+1/2$.

 \item
The related wave functions remain bounded, normalizable and
proportional to the Laguerre polynomials,
 \ben
 \psi^{(reg)}(r) = const \cdot r^{\ell+1}\,e^{-r^2/2}\,
L_n^{\ell+1/2}(r^2),
 \een
\ben
 \psi^{(irreg)}(r) = const \cdot r^{-\ell}\, e^{-r^2/2}\,
L_n^{-\ell-1/2}(r^2).
 \een
They exhibit the same $\alpha \leftrightarrow -\alpha$ symmetry.

 \item
At $g=0$ (i.e., $\ell=0$) we return to the well known
one-dimensional spectrum,
 \ben
 E^{(reg)}_n = 4n+3,
\ \ \ \ \ \ \
 E^{(irreg)}_n =4n+1, \ \ \ \ \ \ \ \ell=0,\ \ \ \ \ \
n=0, 1, \ldots .
 \een
The seemingly formal $\alpha \leftrightarrow -\alpha$ symmetry
degenerates  to the  parity.

\item
The latter feature (parity) survives smoothly the transition to
the singular systems. At all the sufficiently small $g $ the
${\cal PT}$ symmetric regularization preserves the continuity and
reality of the energies as well as their numbering by the integer
$n$ and by the sign of the parameter $\alpha = \ell+1/2
=\sqrt{1/4+g/2}$.

\end{itemize}

 \noindent
The  latter smoothness is in a sharp contrast with the abrupt loss
of the irregular half of the spectrum at any $g \neq 0$ in the
traditional Hermitian formalism. Summarizing the situation, we can
now start from the $g=0$ or $\ell =0$ bound states with the well
defined values of parity $=\pm 1$. A smooth continuation in~$g$
transfers the label $\pm 1$ to all the energies. The complexified
bosonic and fermionic symmetry becomes re-established in this
manner. This enables us to speak about the bosons and fermions
defined by the following rule
 \be
\ba
 \psi^{(bosonic)}(-{r})
= (-1)^{-\ell} \psi^{(bosonic)}({r}), \ \ \ \ \\
\psi^{(fermionic)}(-{r}) = (-1)^{\ell+1}
 \psi^{(fermionic)}({r}).
\ea
 \label{statpt}
 \ee
The new statistics (\ref{statpt}) resembles strongly some aspects
of its Hermitian alternative. In essence, the transition from the
new fermions to bosons is just a change of sign of
$\alpha=\ell+1/2$. Such a ``complexified supersymmetry" differs
from its older form in~\cite{PLB}.



\newpage

 \begin{thebibliography}{00}

\bibitem{Messiah}
Messiah A.: Quantum Mechanics II. North Holland, Amsterdam, 1986.

\bibitem{Witten}
Witten E.: Nucl. Phys. B 188 (1981) 513.

\bibitem{Khare}
Cooper F., Khare A. and Sukhatme U.: Phys. Rep. 251 (1995) 267.

\bibitem{BB}
Bender C. M. and Boettcher S.: Phys. Rev. Lett. { 24} (1998) 5243.

\bibitem{BBjmp}
Bender C. M., Boettcher S. and Meisinger P. N.: J. Math. Phys. 40
(1999) 2201.

\bibitem{PLB}
Znojil M., Cannata F., Bagchi B. and Roychoudhury R.: Phys. Lett.
B 483 (2000) 284.

\bibitem{Reed}
Reed M. and Simon B.: Methods of Modern Mathematical Physics IV,
Academic Press, New York, 1978.

\bibitem{Mandal}
Khare A. and Mandal B. P.: Phys. Lett. A 272 (2000) 53.

\bibitem{Delaba}
Delabaere E. and Trinh D. T.: J. Phys. A: Math. Gen. 33 (2000)
8771.

\bibitem{BM}
Bender C. M. and Milton K. A.: Phys. Rev. D 55 (1997) R3255.

\bibitem{refer}
I am obliged to anonymous referee for his/her constructive
criticism of the insufficient emphasis put on these
``philosophical" quyestions in my MS before revision.

\bibitem{Calogero}
Calogero F.: J. Math. Phys. 10 (1969) 2191.

\bibitem{Darboux}
Darboux G.: Comptes Rendus Acad. Sci. (Paris) 94 (1882) 1456.

\bibitem{Comment}
Znojil M.: Phys. Rev. A 61 (2000) 066101.

\bibitem{Ditt}
Dittrich J. and Exner P.: J. Math. Phys. 26 (1985) 2000.

\bibitem{Caliceti}
Caliceti E., Graffi S. and Maioli M.: Commun. Math. Phys. 75
(1980)
 51.

\bibitem{DB}
Bessis D.: private communication (1992);

Delabaere E. and Pham F.: Phys. Letters A 250 (1998) 25 and 29;

Fern\'andez F. M., Guardiola R., Ros J. and Znojil M.: J. Phys. {
A}: Math. Gen. 31 (1998) 10105;

Andrianov A. A., Ioffe M. V.,  Cannata F. and Dedonder J. P., Int.
J. Mod. Phys. A 14 (1999) 2675;

Mezincescu G. A.: J. Phys. A: Math. Gen. 33 (2000) 4911.

\bibitem{Alvarez}
Alvarez G.: J. Phys. A: Math. Gen. 27 (1995) 4589.

\bibitem{BBjpa}
Bender C. M. and Boettcher S.: J. Phys. A: Math. Gen. 31 (1998)
L273.

\bibitem{dekadic}
Znojil M.: J. Phys. A: Math. Gen. 33 (2000) 6825.

\bibitem{shapein}
Bagchi B. and Roychoudhury R.: J. Phys. A: Math. Gen. 33 (2000)
L1;

Znojil M.: J. Phys. A: Math. Gen. 33 (2000) L61.

\bibitem{Eckart}
Znojil M.: J. Phys. A: Math. Gen. 33 (2000) 4561;

L\'{e}vai G. and Znojil M.: J. Phys. A: Math. Gen. 33 (2000) 7165.

\bibitem{BT}
Bender C. M. and Turbiner A.: Phys. Lett. A 173 (1993) 442.

\bibitem{BG}
Buslaev V. and Grechi V.: J. Phys. A: Math. Gen. 26 (1993) 5541.

\bibitem{hopt}
Znojil M.: Phys. Lett. { A 259} (1999) 220.

\bibitem{quartic}
Znojil M.: J. Phys. A: Math. Gen. 33 (2000) 4203.

\bibitem{Cannata}
Cannata F.,  Junker G. and Trost J.: Phys. Lett. { A 246} (1998)
 219.

\bibitem{Flessas}
G. P. Flessas and R. R. Whitehead, Journal of Mathematical Physics
25 (1984) 910.

\end{thebibliography}
\end{document}